\RequirePackage{fix-cm}
\documentclass[a4paper]{scrartcl}

\usepackage[a4paper]{geometry}
\usepackage[T1]{fontenc}
\usepackage[USenglish]{babel}

\usepackage[babel]{csquotes}
\usepackage{scrhack}
\usepackage{authblk}
\usepackage{floatrow}
\usepackage[backend=biber,bibencoding=utf8,sorting=none,date=year,
style=nature,isbn=false,doi=false,url=false]{biblatex}
\usepackage[onehalfspacing]{setspace}
\usepackage{varioref}
\usepackage[english,plain]{fancyref}
\usepackage[colorlinks,linkcolor=black,urlcolor=black,citecolor=black]{hyperref}
\usepackage[usenames,dvipsnames,svgnames,table]{xcolor}
\usepackage{graphicx} 
\usepackage{tabularx}
\usepackage{multirow}
\usepackage{multicol}
\usepackage{booktabs}
\usepackage{rotating}
\usepackage{float} 
\usepackage{amsmath} 
\usepackage{cleveref}
\Crefformat{figure}{#2Fig.~#1#3}
\usepackage{empheq}
\usepackage{amssymb} 
\usepackage{bm}
\usepackage{commath}
\usepackage{amsfonts} 
\usepackage{mathtools} 
\usepackage[range-phrase={\text{~to~}},output-decimal-marker={.}]{siunitx}
\usepackage[colorinlistoftodos,prependcaption,textsize=tiny]{todonotes}
\usepackage{lipsum}
\usepackage{libertine} 
\usepackage{siunitx}
\usepackage[version=4]{mhchem}
\usepackage[format=plain,figurename=Fig., figurename=Fig.,tablename=Tab.]{caption}
\usepackage{lineno}
\usepackage{soul}

\crefname{figure}{fig.}{figs.}
\Crefname{figure}{Fig.}{Figs.}
\crefname{table}{tab.}{tabs.}
\Crefname{table}{Table}{Tables}
\crefname{equation}{eq.}{eqs.}
\Crefname{equation}{Eq.}{Eqs.}
\DeclareSIUnit\molar{\textsc{M}}
\DeclareSIUnit\photons{\textrm{photons}}
\DeclareSIUnit\pixel{\textrm{pixel}}

\title{Resolving molecular diffusion and aggregation of antibody proteins with megahertz X-ray free-electron laser pulses}

\begin{document}

\author[1]{Mario Reiser\thanks{mario.reiser@fysik.su.se}}
\author[2]{Anita Girelli}
\author[2]{Anastasia Ragulskaya}
\author[1]{Sudipta Das}
\author[1]{Sharon Berkowicz}
\author[1]{Maddalena Bin}
\author[1]{Marjorie Ladd-Parada}
\author[1]{Mariia Filianina}
\author[1,2]{Hanna-Friederike Poggemann}
\author[2]{Nafisa Begam}
\author[3]{Mohammad Sayed Akhundzadeh}
\author[3]{Sonja Timmermann}
\author[3]{Lisa Randolph}
\author[4]{Yuriy Chushkin}
\author[5]{Tilo Seydel}
\author[6]{Ulrike Boesenberg}
\author[6]{J\"org Hallmann}
\author[6]{Johannes M\"oller}
\author[6]{Angel Rodriguez-Fernandez}
\author[6]{Robert Rosca}
\author[6]{Robert Schaffer}
\author[6]{Markus Scholz}
\author[6]{Roman Shayduk}
\author[6]{Alexey Zozulya}
\author[6]{Anders Madsen}
\author[2]{Frank Schreiber}
\author[2]{Fajun Zhang}
\author[1]{Fivos Perakis\thanks{f.perakis@fysik.su.se}}
\author[3]{Christian Gutt\thanks{christian.gutt@uni-siegen.de}}

\affil[1]{Department of Physics, AlbaNova University Center, Stockholm University, S-106 91 Stockholm, Sweden}
\affil[2]{Institut f\"ur Angewandte Physik, Universit\"at T\"ubingen, Auf der Morgenstelle 10, 72076 T\"ubingen, Germany}  
\affil[3]{Department Physik, Universit\"at Siegen, Walter-Flex-Strasse 3, 57072 Siegen, Germany}
\affil[4]{ESRF - The European Synchrotron, 71 Avenue des Martyrs, Grenoble, 38000, France}
\affil[5]{Institut Laue-Langevin, 38042 Grenoble Cedex 9, France}
\affil[6]{European X-Ray Free-Electron Laser Facility, Holzkoppel 4,22869 Schenefeld, Germany}

\date{}
\maketitle

\section*{Abstract}
X-ray free-electron lasers~(XFELs) with megahertz repetition rate can provide 
novel insights into structural dynamics of biological macromolecule solutions.
However, very high dose rates can lead to 
beam-induced dynamics and structural changes due to radiation damage.
Here, we probe the dynamics of dense antibody protein (Ig-PEG) 
solutions using megahertz X-ray photon correlation
spectroscopy (MHz-XPCS) at the European XFEL.
By varying the total dose and dose rate,
we identify a regime for measuring the motion of proteins
in their first coordination shell, 
quantify XFEL-induced effects such as driven motion, and map out the 
extent of agglomeration dynamics. 
The results indicate that for average dose rates below
\SI{1.06}{\kilo\gray\per\micro\second} in a time window up to \SI{10}{\micro\second}, 
it is possible to capture the protein dynamics before the onset of 
beam induced aggregation. We refer to this approach as
\emph{correlation before aggregation} and demonstrate that MHz-XPCS bridges an
important spatio-temporal gap in measurement techniques for biological samples.

\subsection*{Introduction}
The European X-ray Free-Electron Laser Facility~(EuXFEL) is the first X-ray
free electron laser (XFEL) generating ultrashort hard X-ray pulses with
megahertz repetition rate. 
Megahertz X-ray photon correlation spectroscopy~(MHz-XPCS) 
\cite{lehmkuhler_emergence_2020,dallari_microsecond_2021,dallari_analysis_2021} 
makes use of this high repetition rate and the high degree of transverse 
coherence to measure diffusive
dynamics with (sub-) microsecond temporal resolution.
In biological systems, typical diffusion coefficients in dense cellular 
environments range from
$D_0\approx\SIrange{0.1}{10}{\nano\metre\squared\per\micro\second}$ 
\cite{garcia_de_la_torre_calculation_2000,ridgway_coarse-grained_2008,grimaldo_dynamics_2019,roosen-runge_protein_2011,von_bulow_dynamic_2019,grimaldo_protein_2019}
which requires to resolve time scales from 
$\tau\approx\SIrange{0.5}{5}{\micro\second}$ (\Cref{fig:experiment})
in order to trace the complex many-body interactions between proteins 
and the solvent on molecular length scales.
This window of length and time scales is not accessible by optical techniques 
such as dynamic light scattering, which measures longer length scales 
(micrometers), or neutron spectroscopy techniques such as neutron spin echo or
inelastic neutron scattering, which typically measure on faster time scales of nanoseconds 
and below. Clearly, experimental techniques are needed to close this gap and 
measure collective dynamics on microsecond time scales and nanometer length 
scales. By analyzing fluctuating X-ray speckle patterns, MHz-XPCS is potentially
capable of closing this gap, as we demonstrate here, and
enables us to gain information on equilibrium and out-of-equilibrium 
collective dynamics of protein solutions. 

Protein dynamics in crowded environments are particularly relevant in the context
of intracellular transport in the cytoplasm of eukaryotic cells 
\cite{leeman_proteins_2018},
phase transitions in biomolecular condensates 
\cite{myung_weak_2018,bucciarelli_dramatic_2016,tang_protein_2019},
aggregation phenomena \cite{pease_determination_2008,martin_prevention_2014}
and drug production \cite{skar-gislinge_colloid_2019}.
In highly concentrated environments,
the dynamics differ significantly from that of a dilute system,
whereas the exact mechanisms that influence the dynamics on different
time scales are not yet fully understood 
\cite{roosen-runge_protein_2011,grimaldo_dynamics_2019,girelli_molecular_2021}.
It was found that \emph{in vivo} dynamics in cells exhibit tremendously reduced
diffusion compared to \emph{in vitro} measurements of diluted proteins in buffer
solutions 
\cite{wang_effects_2010,li_translational_2009,london_nuclear_1975,williams_19f_1997,ando_crowding_2010,wojcieszyn_diffusion_1981,arrio-dupont_translational_2000,verkman_solute_2002}.
It is believed that the level of slowing-down depends on the particular system 
and possibly additional crowding agents 
\cite{banks_anomalous_2005,wang_effects_2010,muramatsu_tracer_1988,dix_crowding_2008,grimaldo_dynamics_2019}.
In addition to excluded volume effects 
\cite{zimmerman_macromolecular_1993,mukherjee_macromolecular_2015},
there can be contributions from the local
water dynamics of the hydration layer \cite{harada_protein_2012}, quinary
interactions of proteins with other cytoplasmic
constituents \cite{cohen_cell_2017,mcconkey_molecular_1982,yu_biomolecular_2016,wang_effects_2010,li_translational_2009,feig_variable_2012},
and transient cluster formation 
\cite{cardinaux_cluster-driven_2011,kowalczyk_equilibrium_2011,porcar_formation_2010,cardinaux_cluster-driven_2011,liu_lysozyme_2011,nawrocki_slow-down_2017}
that influence intracellular protein diffusion.
Also, the dynamics often exhibit anomalous behavior--i.e., non-Brownian
and in particular subdiffusive dynamics 
\cite{banks_anomalous_2005,weiss_anomalous_2004,guigas_sampling_2008}--and 
making it difficult to extrapolate the dynamics from the dilute regime.
Clearly, new methods are needed to directly probe diffusive dynamics in crowded 
biological solutions on (sub-) microsecond time scales and nanometer 
length scales to study these phenomena.

Radiation damage constitutes a major challenge for X-ray scattering experiments
with protein solutions.
Radiolysis of water and the fast distribution
of the free radicals formed rapidly degrade the protein molecules. Hence,  
a typical upper limit of tolerable absorbed doses is estimated on the order
of a few kilo Gray in these experiments with the exact value  depending
on the chemical composition of the system 
\cite{meisburger_breaking_2013,hopkins_quantifying_2016,garman_x-ray_2017,kuwamoto_radiation_2004}. 
Protein aggregation is a signature of beam-induced damage in protein
solutions visible via changes in the X-ray scattering form factor.
Aggregation processes and the spread of free radicals are both driven by
diffusive dynamics and act on nano- and microsecond time scales 
\cite{pease_determination_2008,young_photon-photon-out_2021,hawkins_generation_2001,khalack_solvation_2005}. 
The study of such time-dependent dynamic processes in aqueous solutions
of bio-molecules when illuminated with X-rays 
is of considerable relevance for understanding biological aspects of ionizing
radiation.
In addition, MHz-XFEL experiments deliver extremely high dose rates to the sample. 
Utilizing MHz repetition rates and high attenuation,
the X-ray pulses are delivered on (sub-)~microsecond time scales
such that an average dose rate on the order of several kilo Gray per microsecond can be reached.
The effects of such high dose rates on structure and dynamics of protein solutions are still unknown. 

Here, we report a MHz-XPCS experiment with radiation sensitive
protein solutions at the Materials Imaging and Dynamics~(MID)
instrument \cite{madsen_materials_2021} at EuXFEL. 
We investigate the dynamics in a concentrated bovine immunoglobulin~(Ig)
solution where \SI{80}{\percent} of the Ig is constituted by
IgG \cite{da_vela_effective_2017,girelli_microscopic_2021}.
Immunoglobulin is an abundant antibody protein that can be found, for instance,
in the blood of animals and humans.
Polyethylene glycol~(PEG) is added to the solution as a
depletant and induces attractive protein-protein interactions that--depending
on concentration and temperature--can result in liquid-liquid phase
separation~(LLPS) \cite{da_vela_effective_2017,girelli_microscopic_2021}.
This combination renders the Ig-PEG system an interesting candidate for the
MHz-XPCS measurements in the context of both crowding dynamics in concentrated
protein solutions and the formation of biomolecular condensates.

\section*{Results}
\subsection*{Measurement Scheme And Data Collection}
\begin{figure}[H]
  \centering
  \includegraphics[width=\linewidth]{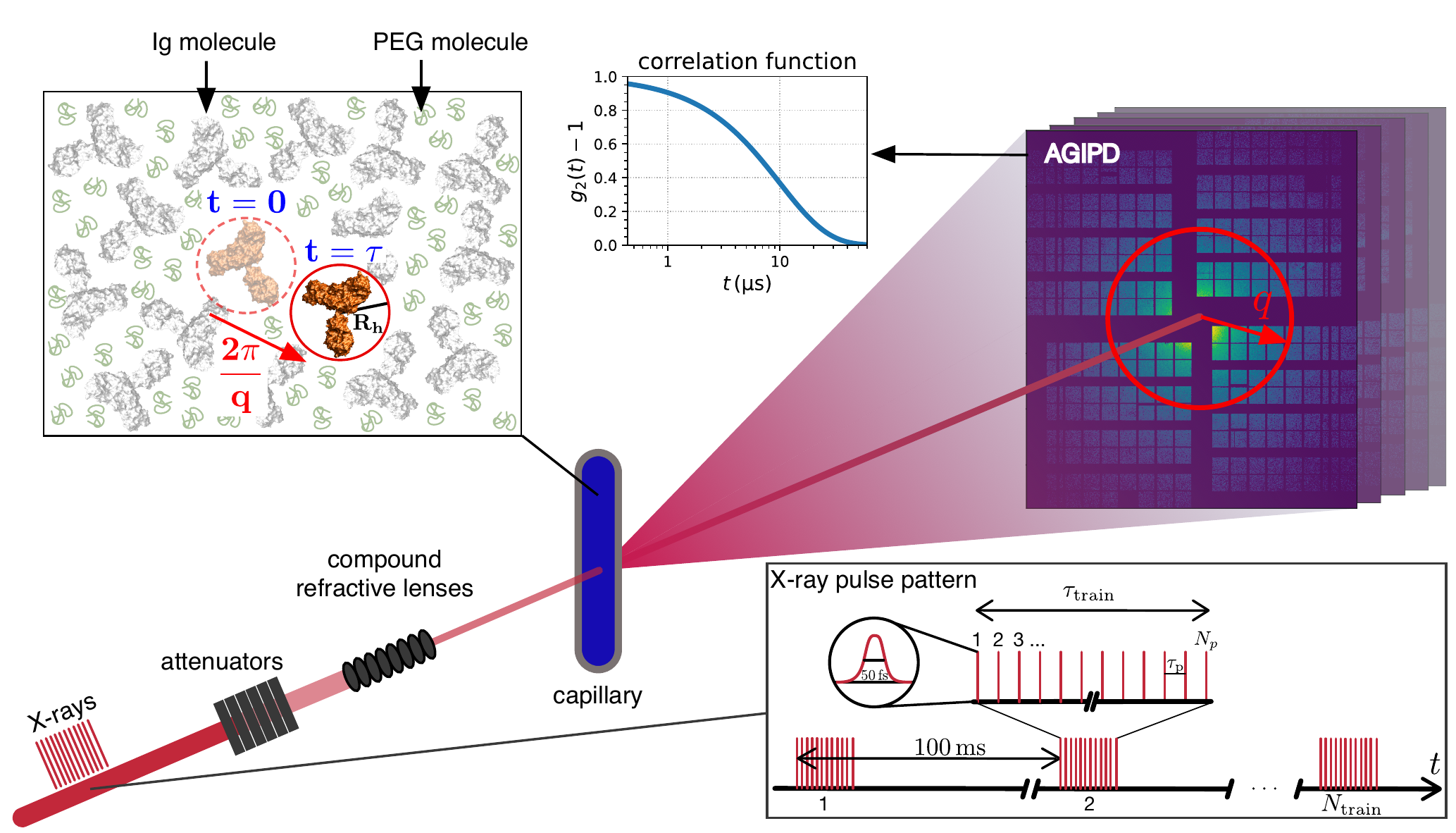}
  \caption[Experiment]{\textbf{Scheme of the experiment.}
  Highly concentrated solutions of
    immunoglobulin (Ig) with polyethylene glycol~(PEG) are measured in quartz 
    capillaries. 
    An individual Ig molecule has a hydrodynamic radius of $R_h=\SI{5.5}{\nm}$.
    Megahertz X-ray photon correlation spectroscopy (MHz-XPCS)
    measurements are performed by using trains of X-ray
    pulses, which illuminate the sample. The spacing between two pulses within a
    train is $\tau_{p}$ and was varied between \SIlist{443;886}{\ns}
    where a train contains $N_p$ individual X-ray pulses.
    The length of an individual X-ray pulse in the time domain is $\leq\SI{50}{\fs}$~\cite{madsen_materials_2021}. A new train is delivered every \SI{100}{\ms}.
    The train duration is determined by the number of pulses per train and the delay 
    time between the pulses: $\tau_{\mathrm{train}}=(N_p-1)\tau_p$.
    The longest train duration during the experiment
    was $\tau_{\mathrm{train},max}=(144-1) \times \SI{886}{\ns}\approx\SI{127}{\micro\s}$.
    For a period of $\SI{100}{\ms}-\tau_{\mathrm{train}}$, the sample is not illuminated by X-rays.
    By analyzing sequential X-ray scattering patterns measured with 
    the adaptive gain integrated pixel detector (AGIPD), information about the dynamics
    of the sample can be obtained in the form of intensity auto-correlation
    functions calculated from fluctuating speckle patterns. A measurement consists 
    of a series of $N_\mathrm{train}$ invdividual trains (\Cref{tab:beamline-params}).
  }\label{fig:experiment}
\end{figure}

We employed X-ray pulses with \SIlist{443; 886}{\ns} delays between successive pulses corresponding roughly to repetition rates of \SIlist{2.26;1.13}{\mega\hertz}, respectively.
The X-ray pulses were delivered in trains of up to \num{200} pulses with a train
frequency of \SI{10}{\hertz} (\Cref{fig:experiment}).
This time structure makes it possible to conduct MHz-XPCS measurements 
within a single train, while the time between 
subsequent trains is 
sufficiently long to refresh the sample via translation.

The data presented here were acquired at the MID instrument in small-angle X-ray
scattering~(SAXS) geometry with a pink beam, i.e., using self-amplified 
spontaneous emission~(SASE) without a monochromator,
and a photon energy of \SI{9}{\kilo\electronvolt} \cite{madsen_materials_2021}.
A sketch of the experimental setup is shown in \Cref{fig:experiment}.
The Adaptive Gain Integrating Pixel Detector~(AGIPD) 
\cite{allahgholi_adaptive_2019} was placed \SI{7.46}{\m} behind the sample with
most of the sample-detector flight path being evacuated.
The Ig-PEG solutions were filled into quartz capillaries with an outer diameter 
of \SI{1.5}{\mm} and a wall thickness of \SI{20}{\micro\metre}.
A Linkam scientific instruments stage was used to control and stabilize the 
sample temperature at \SI{298}{\kelvin}, which is above the binodal 
in the single phase regime of the Ig-PEG system
\cite{girelli_microscopic_2021,da_vela_effective_2017}.
The X-ray beam was focused to a diameter of ${\SI{10}{\mu\m}}$ (FWHM) using compound
refractive lenses to increase the measured speckle contrast and the
signal-to-noise~(SNR) of the XPCS measurements \cite{falus_optimizing_2006}.

\Cref{tab:beamline-params} contains a summary of the measurement parameters.
The intensity of the X-rays was reduced by chemically vapour deposited~(CVD)
diamond attenuators of various thickness and adjusted such that
the samples were exposed to the lowest possible dose while keeping the scattered
intensity high enough to reach a sufficient SNR.
For example, with an average pulse energy of \SI{1.2}{\m\J} and
${\SI{3925}{\mu\m}}$ CVD attenuator thickness \SI{6.5e8}{\photons} per 
X-ray pulse illuminate the sample. The incoming flux results in an average
scattering signal of less than \SI{e-1}{\photons} per pixel per image. 
In addition to the absolute dose also the average dose rate was varied, i.e.,
the absorbed dose per time, measured in \si{\kilo\gray\per\micro\second}. 
The actual dose rate value is calculated as an average over the first ten 
X-ray pulses and all trains of a measurement (see Methods).

\subsection*{Megahertz small angle X-ray scattering (MHz-SAXS)}
\begin{figure}[H]
  \centering
  \includegraphics[width=\linewidth, trim={0 4mm 0 0}]{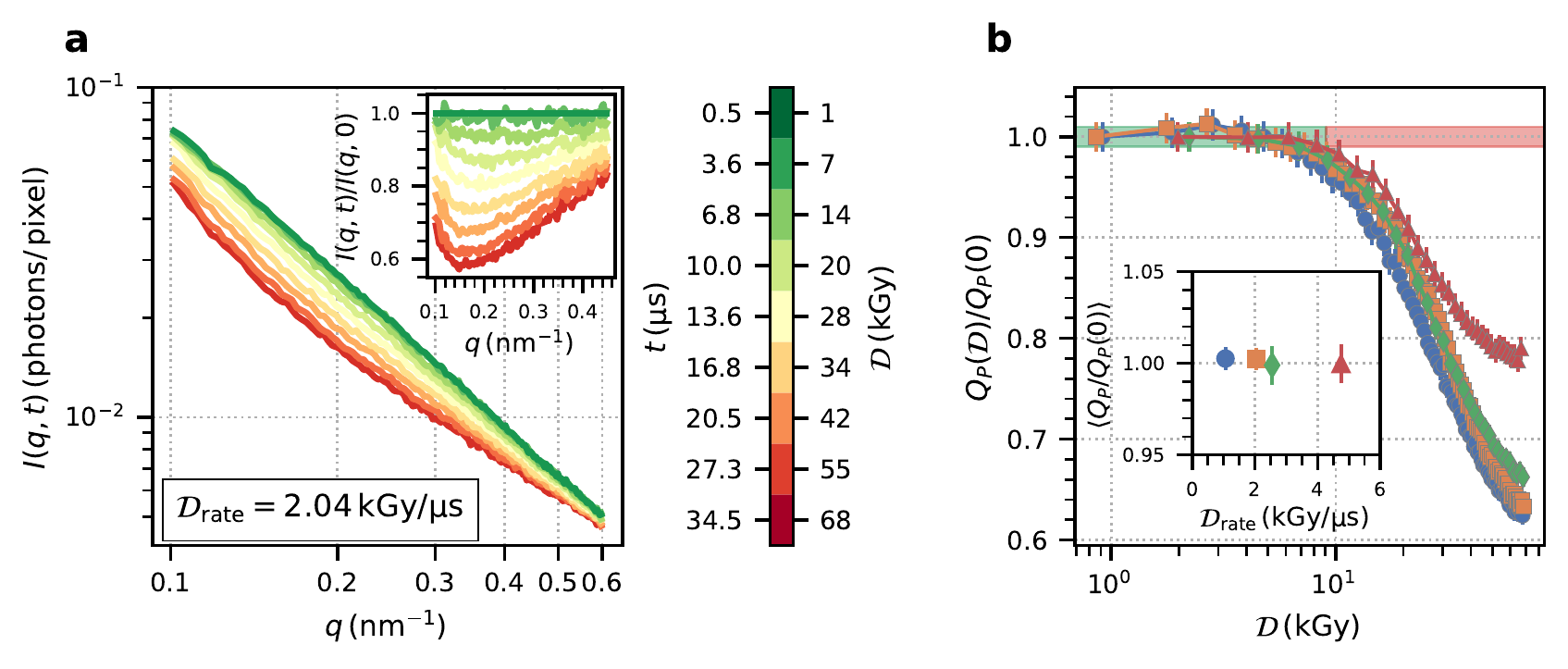}
  \caption[structural-changes]{\textbf{Static scattering signal of Ig-PEG.}
    \textbf{a}~The azimuthally integrated intensity, ${I(q, t)}$, as a function
    of momentum transfer, $q$. The color indicates the absorbed dose
    $\mathcal{D}$ and the corresponding timescales. 
    The data shown are acquired with a
    dose rate of \SI{2.04}{\kilo\gray\per\micro\second}.
    The inset displays $I(q, t)$ normalized to the first pulse $I(q, 0)$. 
    \textbf{b}~Porod invariant, ${Q_P(\mathcal{D})}$, calculated from the data displayed in~\textbf{a} (orange) and three additional dose rates.
    The data are normalized to ${Q_P(0)}$ and the error bars are calculated as the standard deviation
    of the normalized second pulse from unity. The inset shows the mean
    of ${Q_P(\mathcal{D})}$ below \SI{10}{\kilo\gray} for different dose
    rates. The error bars indicate the weighted standard deviation of $Q_P$
    in this range. Source data are provided as a Source Data file.}\label{fig:structural-changes}
\end{figure}

The evolution of the time-resolved SAXS signal as a function of dose and dose
rate is analyzed by computing the azimuthally integrated intensity $I(q,t)$
as a function of absolute momentum transfer, ${q=4\pi/\lambda\sin{(2\theta/2)}}$,
where $\lambda$ is the X-ray wavelength and $2\theta$ is the scattering angle,
and measurement time or dose (\Cref{fig:structural-changes}a).
The absorbed dose is proportional to the measurement time
and is calculated with \Cref{eq:radiation-dose} (see Methods).
The data displayed were recorded with a dose rate of \SI{2.04}{\kilo\gray\per\micro\second},
but with absolute doses varying between \SI{1}{\kilo\gray}~(green) and \SI{74}{\kilo\gray}~(red).
With increasing dose, we observe significant changes in $I(q,t)$, with the largest
decrease of intensity visible at momentum transfers of ${q=\SI{0.17}{\per\nm}}$
accompanied by an increasing scattering signal at small momentum transfers.
The inset shows the data normalized by $I(q,0)$.
The Ig-PEG system exhibits a structure factor peak close to \SI{0.62}{\per\nm}
that was studied in a previous work by \textcite{da_vela_effective_2017}.
Additionally, the phase behavior of the Ig-PEG systems is characterized by an
upper critical solution temperature of around \SI{294}{\kelvin}. 
The overall decrease of intensity in \Cref{fig:structural-changes} indicates
that the system is moving away from the 
LLPS binodal in the phase diagram, presumably due to beam-induced local heating.
Deeper in the single phase regime, increasingly repulsive protein-protein
interactions lead to a reduced SAXS intensity.
On the other hand, at $q$-values below \SI{0.17}{\per\nm}, the visible
increase of $I(q,t)/I(q,0)$ indicates the formation of X-ray induced aggregation
of the proteins.

We quantify the evolution of structural changes by calculating
the Porod invariant
\begin{equation}
  \label{eq:porod-invariant}
    Q_P(t) = \int_{q_\mathrm{\min}}^{q_\mathrm{\max}} q^2 I(q, t) \,dq\,,
\end{equation} 
in the accessible $q$-range
(${q_\mathrm{min}=\SI{0.1}{\per\nm}}$, ${q_\mathrm{max}=\SI{0.6}{\per\nm}}$) as a function of dose
(\Cref{fig:structural-changes}b). 
$Q_P(\mathcal{D})$ displays an initial plateau up to a maximum dose of \SI{10}{\kilo\gray}
after which it starts to decrease more than two percent from its initial value.
At doses below \SI{10}{\kilo\gray}, the protein structure seems unaffected 
by the X-ray illumination--at least on the length scales probed here.
In this low-dose regime, we also extract the dose rate dependence of
the Porod invariant by averaging the $Q_p(\mathcal{D})$
data for  ${\mathcal{D}<\SI{10}{\kilo\gray}}$.
The results are displayed in the inset in \Cref{fig:structural-changes}b 
demonstrating the absence of a dose rate dependence in the SAXS signal.
This is in agreement with previous work reporting that the absolute 
absorbed dose is the main driver for radiation damage and dose rate effects
are only weak \cite{kuwamoto_radiation_2004}.

\subsection*{Megahertz X-ray photon correlation spectroscopy (MHz-XPCS)}
The disordered protein solutions give rise to a speckle pattern
in the far-field when illuminated by coherent radiation.
The dynamics can be studied by analyzing the speckle intensity fluctuations
that are related to the microscopic motion of the protein molecules.
The intensity ${I_p(q, t)}$ is measured at time $t$ by pixel $p$ within
a concentric region of interest~(ROI) of constant absolute momentum transfer.
We utilized an XPCS adapted acquisition scheme in which the sample is 
continuously moving through the X-ray beam with \SI{400}{\micro\metre\per\second}.
The sample movement is negligible during an X-ray train ensuring
illumination of the same sample spot on microsecond time scales.
In between two trains the sample position offset is large enough to completely 
renew the sample volume, and thus to avoid accumulated damage.
The low intensity scattering signal requires averaging correlation functions
from many trains (between \num{2000} and \num{9000}  
\Cref{tab:beamline-params}) to increase the SNR. 
Approximately  \SI{80}{\percent} of the acquired trains are used for the XPCS analysis
while the rest are discarded after applying filters based on diagnostics such as
extremely low intensity due to the SASE fluctuations.

We compare measurements with average dose rates, ${\mathcal{D}_\mathrm{rate}}$,
from \SIrange{1.06}{4.75}{\kilo\gray\per\micro\second}.
The influence of dose and dose rate on the protein dynamics can be quantified
with the help of two-time correlation functions~(TTCs) 
\cite{sutton_using_2003,cipelletti_time-resolved_2003} 
which essentially represent the correlation coefficient between speckle images
taken at times $t_1$ and $t_2$ at momentum transfer $q$:
\begin{equation}
  \label{eq:two-time-correlation}
  c_2(q,t_1,t_2)=\bigg\langle\frac{\langle I_p(q,t_1)I_p(q,t_2)\rangle_p}{\langle
    I_p(q,t_1)\rangle_p\langle I_p(q,t_2)\rangle_p}\bigg\rangle_j-1\,.
\end{equation}
Here, ${\langle \dots \rangle_p}$ denotes an average over all pixels 
with the same absolute momentum transfer, $q$, and ${\langle \dots \rangle_j}$ denotes an average over all trains where $j=1\dots N_\mathrm{train}$. The data calibration and
analysis workflow for MHz-XPCS with AGIPD is described in detail in \textcite{dallari_analysis_2021}.

\begin{figure}[htbp]
  \centering
  \includegraphics[width=\linewidth, trim={0 4mm 0 0}]{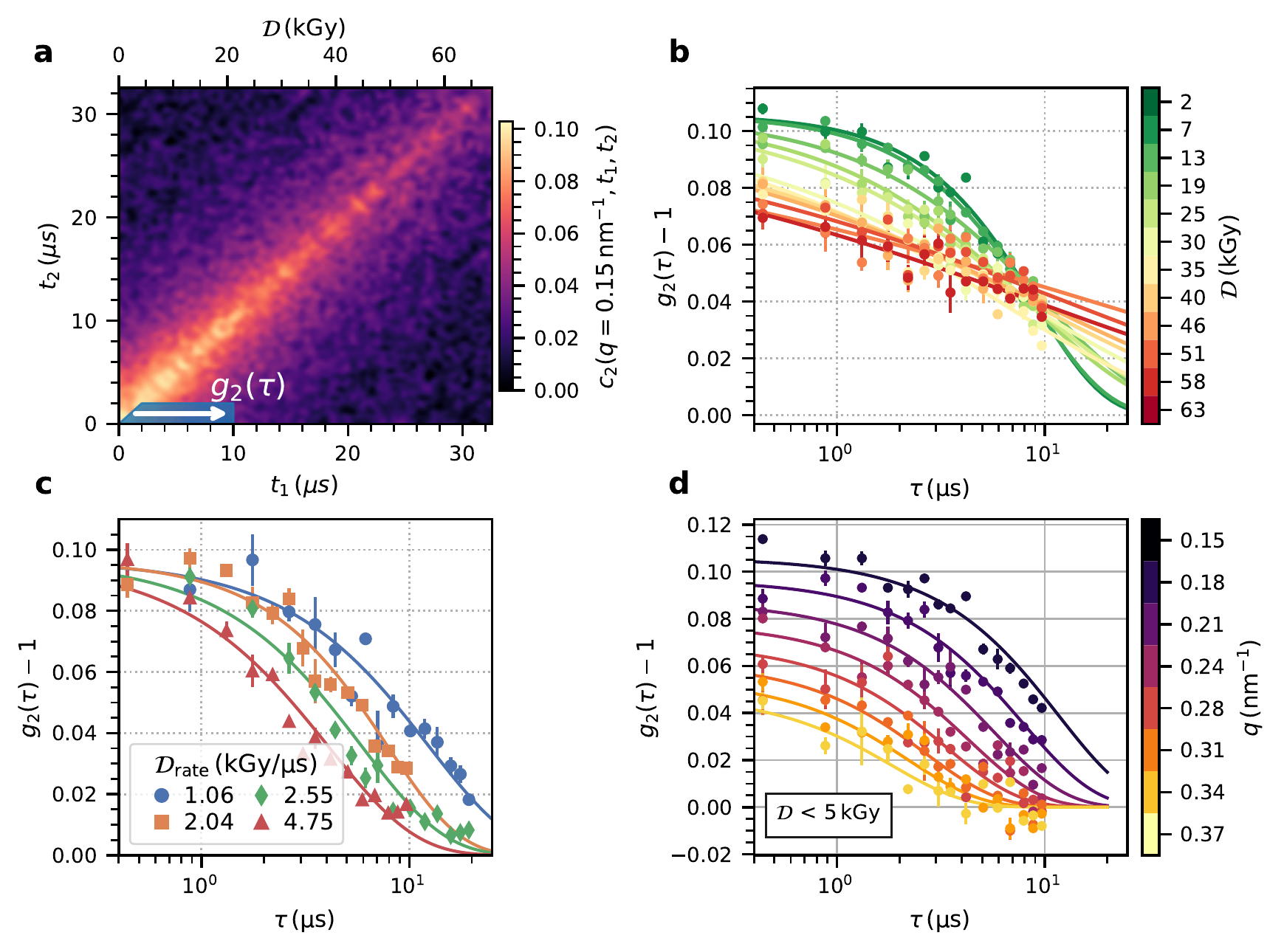}
  \caption[ttc-4panels]{\textbf{Correlation functions.} \textbf{a}~Two-time correlation function, $c_2$, of
    Ig-PEG measured with an average dose rate of
    $\SI{2.04}{\kilo\gray\per\micro\second}$ for
    ${q=\SI{0.15}{\per\nano\metre}}$. \textbf{b}~Correlation functions for
    different initial doses 
    ($\mathcal{D}_{\mathrm{rate}}=\SI{2.04}{\kilo\gray\per\micro\second}$, 
    ${q=\SI{0.15}{\per\nano\metre}}$). 
    \textbf{c}~Correlation functions with an initial dose below
    \SI{5}{\kilo\gray} for different dose rates at ${q=\SI{0.15}{\per\nano\metre}}$.
    \textbf{d}~correlation functions for different momentum transfers 
    fitted with a $q$-squared dependent relaxation rate 
    ($\mathcal{D}_\mathrm{rate}=\SI{2.04}{\kilo\gray\per\micro\second}$). The error bars represent the standard error over pixels and repetitions. Source data are provided as a Source Data file.
    }\label{fig:correlation-functions}
\end{figure}

\Cref{fig:correlation-functions}a displays a TTC measured with
an average dose rate of
$\SI{2.04}{\kilo\gray\per\micro\second}$ at ${q=\SI{0.15}{\per\nm}}$. 
The abscissa and ordinate of the TTC show the measurement times $t_1$ and $t_2$,
respectively, within an X-ray pulse train while the additional label
at the top indicates the corresponding absorbed dose.
The TTC decays with increasing distance from the diagonal describing the
temporal decorrelation of the speckle fluctuations due to the sample dynamics.
The fact that the diagonal does not exhibit a constant width indicates
that the dynamics change throughout the measurement.

Time-resolved intra-train intensity auto-correlation 
functions,  ${g_2(q,\tau, \mathcal{D}(t_0))}$,
are calculated by averaging sections of the TTCs as indicated by the white arrow 
in \Cref{fig:correlation-functions}a:
\begin{equation}
  \label{eq:time-resolved-correlation}
  g_2(q,\tau, \mathcal{D}(t_0))=\langle c_2(q,t_1=t_0+\tau,t_2=t_0\pm \Delta t)\rangle_{\Delta t}+1\,.
\end{equation}
To obtain the correlation function for a particular
initial dose,
$\mathcal{D}(t_0)$, during the measurement, the time $t_0$ is chosen on the diagonal of the TTC in \Cref{fig:correlation-functions}a,  while noting that the dose increases further with each point 
of the correlation function.
The average over $\Delta t$ can be seen as rebinning $c_2$ along $t_2$ to increase the statistics. This approach yields a set of 
$g_2(q,\tau)$ per dose and dose rate.

The correlation functions are modelled by a Kohlrausch-Williams-Watts~(KWW)
function:
\begin{equation}
  \label{eq:g2-fit}
  g_2(q, \tau) = 1 +\beta(q)\,e^{-2(\Gamma(q)\tau)^\alpha}\,,
\end{equation}
where $\beta(q)$ is the $q$-dependent speckle contrast~\cite{hruszkewycz_high_2012} 
($\beta(q=\SI{0.15}{\per\nm}) \approx \SI{11}{\percent}$) and 
$\alpha$ is the KWW exponent.
Brownian diffusion is characterized by a quadratic $q$-dependence of the 
relaxation rates $\Gamma(q)=D_0q^2$, where $D_0$ is the diffusion coefficient,
and simple exponential behavior (${\alpha=1}$).
KWW exponents smaller than one are typically observed in supercooled liquids, 
and gels and can indicate heterogeneous dynamics with a distribution of 
relaxation times \cite{begam_kinetics_2021}.
A quadratic $q$-dependence and a $q$-independent KWW exponent are used to model
the data.
\Cref{fig:correlation-functions}d shows that within the experimental
accuracy the KWW function describes the data well.

\begin{figure}[H]
  \centering
  \includegraphics[width=\linewidth, trim={0 4mm 0 0}]{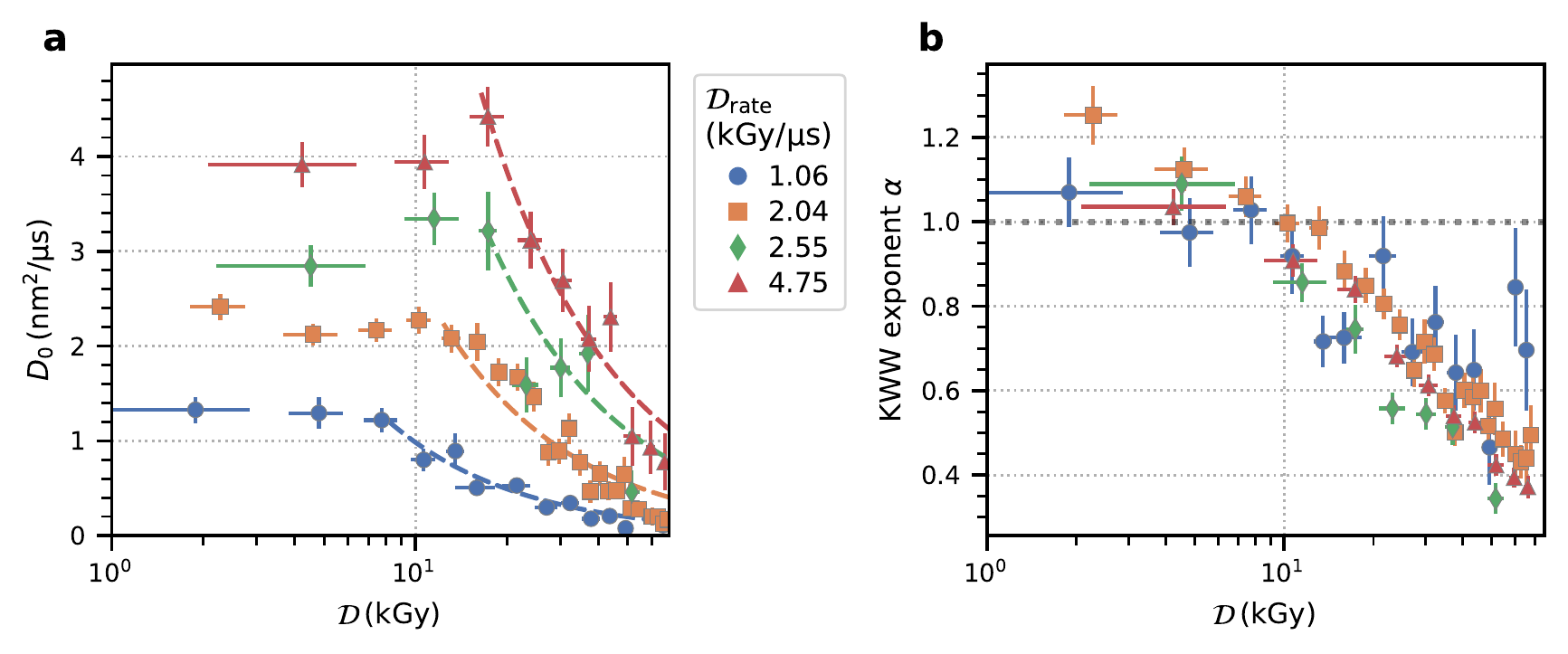}
  \caption[Dstar-vs-dose]{\textbf{Dynamical parameters.} \textbf{a}~Diffusion coefficient,
    $D_0$, as a function of initial dose for different dose rates indicated by the color.
    The dashed lines are guides to the eye.
    \textbf{b}~KWW exponent for different dose rates as a function of total
    absorbed dose.  Source data are provided as a Source Data file.
    }\label{fig:diffusion-coeff-kww-exponent}
\end{figure}

\Cref{fig:correlation-functions}c shows correlation functions for different dose
rates for absolute doses below \SI{5}{\kilo\gray}.
\Cref{fig:correlation-functions}c indicates that the dynamics become 
faster with dose rate as the correlation functions shift to shorter time scales 
while the overall lineshape appears to change only slightly. 
This is different from the behavior observed with increasing total dose in 
\Cref{fig:correlation-functions}b where the shape of the correlation functions 
drastically changes from a simple exponential decay at low doses to a highly
stretched ($\alpha < 1$) and almost logarithmic decay at higher dose values.
We account for these changing KWW exponents by computing the average relaxation 
rate \cite{ruta_wave-vector_2020,guo_entanglement-controlled_2012}
${\langle \Gamma\rangle(q) = \Gamma(q)\alpha / \Gamma_f(1/\alpha)}$, 
where $\Gamma_f(x)$ is the $\Gamma$-function.
Using these average relaxation rates one pair of parameters ${(D_0,\, \alpha)}$
is calculated per initial dose and dose rate, where ${D_0=\langle\Gamma\rangle(q)/q^2}$.
The results are displayed in \Cref{fig:diffusion-coeff-kww-exponent}.

The diffusion coefficients in \Cref{fig:diffusion-coeff-kww-exponent}a reveal a  
pronounced dependence on the initial dose and dose rate
as already indicated 
by the correlation functions in \Cref{fig:correlation-functions} and are higher than expected for the base temperature of $T_0=\SI{298}{\kelvin}$. Therefore,
we denote $D_0$ reported here is as an effective
diffusion coefficient discussed in more detail in the following section.
The numbers obtained for $D_0$ are on the order of a few \si{\nm\squared\per\micro\second}, which is the typical range of 
diffusion coefficients found for dense protein
systems \cite{grimaldo_dynamics_2019,banks_anomalous_2005,roosen-runge_protein_2011}.
For a given dose rate, all diffusion coefficients follow a similar pattern as a
function of initially absorbed dose: $D_0$ is nearly independent of the initial dose up to a 
threshold value, above which the $D_0$ steadily decreases.
The threshold initial dose is slightly below \SI{10}{\kilo\gray} 
for a dose rate of \SI{1.06}{\kilo\gray\per\micro\second}
and increases above \SI{15}{\kilo\gray} for \SI{4.75}{\kilo\gray\per\micro\second}.
The average values of $D_0$ below these thresholds increase linearly
by about a factor of four from \SIrange{1.3}{4}{\nm\squared\per\micro\second}
with increasing dose rate. 
The corresponding KWW exponents do not show any pronounced dose rate dependence,
but a clear dependence on the initial dose (\Cref{fig:diffusion-coeff-kww-exponent}b).
The KWW exponents further reveal that the correlation functions exhibit a simple
exponential shape ($\alpha$=1) for low doses while they are increasingly 
stretched above \SI{10}{\kilo\gray}, which approximately coincides with 
the dose value where a decrease in $D_0$ becomes apparent 
(\Cref{fig:diffusion-coeff-kww-exponent}a).
The simultaneous decrease of $D_0$ and the KWW exponent for high doses points
towards beam-induced
aggregation of the proteins (cf.~\Cref{fig:structural-changes}),
which results in slower diffusion and increasingly stretched exponential
behavior.  

\section*{Discussion}
Our results indicate that static and dynamic properties are influenced
in different ways by the intense X-ray pulses of the European XFEL.
MHz-SAXS reveals that the static scattering signal--within the accessible 
$q$-window--is preserved below an absorbed dose of \SI{10}{\kilo\gray}. 
This threshold value is independent of the applied dose rate
(\Cref{fig:structural-changes}b inset) within the limited range of dose rates.
It is noteworthy that the extremely high dose rates and microsecond time scales
probed with an XFEL yield similar threshold values ($\approx\SI{10}{\kilo\gray}$)
as the orders of magnitude lower dose rates used at a synchrotron 
($\approx\SI{1}{\kilo\gray\per\second}$ 
\cite{girelli_microscopic_2021}). 

Understanding the dose rate dependence of radiation-induced effects is crucial for 
comparing and optimizing experiments at different radiation sources (rotating anodes, 
synchrotrons, XFELs). At comparably moderate dose rates of tens of Gray per second 
at synchrotrons, the aggregation rate of proteins was found to exhibit a dose rate
dependence~\cite{kuwamoto_radiation_2004} favoring measurements with low dose rates. 
On the other hand, high dose rates seem to be preferable in room temperature protein 
crystallography 
measurements~\cite{de_la_mora_radiation_2020,southworth-davies_observation_2007,warkentin_lifetimes_2017}.  


Generally, radiation damage in aqueous protein solutions is mainly attributed
to the diffusion and successive reaction of proteins with radicals produced by 
radiolysis, such as \ce{OH-}.
Radiolysis itself involves a variety of different time and length scales where
the radicals are not uniformly generated in the solvent, but distributed 
initially in nanoscale traces which broaden and diffuse into the bulk on
timescales of hundreds of nanoseconds to microseconds during the chemical stage 
\cite{schwarz_applications_1969}.
The primary yield of \ce{OH-} radicals is high, with \num{2.87} \ce{OH-} per 
\SI{100}{\electronvolt} absorbed after one microsecond
\cite{buxton1987radiation}, leading to an average of about \num{0.6} \ce{OH-}
radicals per Ig protein molecule needed to induce measurable changes
to the SAXS signal (\Cref{fig:structural-changes}).
The observed threshold dose of \SI{10}{\kilo\gray} represents a typical 
time window of \SI{9.4}{\micro\second} when using a dose rate of
\SI{1.06}{\kilo\gray\per\micro\second}.
The absence of a measurable dose rate effect on this static threshold value
indicates that diffusion rates of radicals, recombination and quenching effects
do not affect the overall agglomeration probability. 

Additional insight can be obtained from the MHz-XPCS data, 
which allows to trace time-resolved non-equilibrium dynamics via the TTCs.
With regard to protein diffusion the typical mean square distances
probed here can be estimated via 
\begin{equation*}
    g_2(q,\tau)-1=\beta \exp(-q^2 \langle\Delta x^2(\tau)\rangle /6) 
\end{equation*}
yielding values of  $\sqrt{\langle\Delta x^2\rangle} = \SI{16}{\nm}$ at
$q=\SI{0.15}{\per\nm}$ to $\sqrt{\langle\Delta x^2\rangle} =\SI{4}{\nm}$  at
$q=\SI{0.6}{\per\nm}$ at $1/e$ decay of the correlation functions.
Thus, in the present configuration, MHz-XPCS is sensitive to the motion inside
the first coordination shell of the protein molecules in the dense solution.

Furthermore, it is interesting to examine the dynamics at dose values below 
the static damage threshold of
\SI{10}{\kilo\gray} obtained from the MHz-SAXS analysis.
The diffusion constants are almost independent of the initial
dose  for a given dose rate.
However, $D_0$ displays a pronounced rate dependence and increases
by almost a factor of four between the lowest and the highest dose rate
(\Cref{fig:diffusion-coeff-kww-exponent}b).
Illuminating a sample with highly intense X-ray pulses can lead to a temperature 
increase. Based on the X-ray beam size of \SI{10}{\micro\metre},
we estimate that the generated heat dissipates with a time constant of
\SI{310}{\micro\s} (see Methods section below),
which is much longer than the measurement window
covered by a $g_2$-function here (\SI{20}{\micro\second}).
Thus, the illuminated sample volume does not cool down noticeably
during a measurement and the maximum accumulated heat only depends
on the fluence per pulse and the number of pulses illuminating
the same sample volume which is equivalent to the accumulated dose.

\begin{figure}[H]
  \centering
  \includegraphics[width=\linewidth, trim={0 4mm 0 0}]{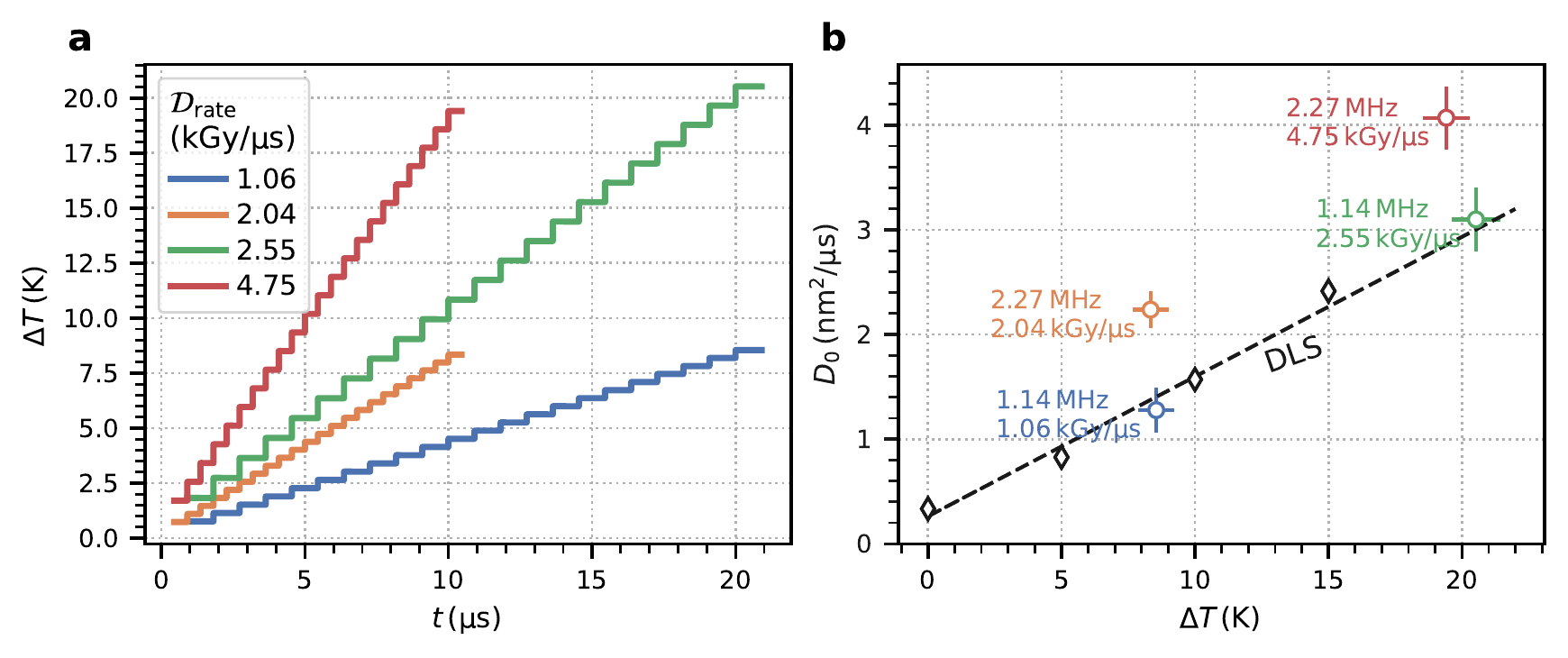}
  \caption[temperature-dose-rate]{\textbf{Effect of temperature on the dynamics.}
  \textbf{a}~Calculated 
    temperature increase $\Delta T$ 
    after \num{20} XFEL pulses for the four different dose rates.  
    \textbf{b}~Diamonds: Temperature dependence of diffusion coefficients measured by dynamic light scattering (DLS). 
    The dashed black line is a linear fit to the data. 
    Circles: diffusion constants determined via XPCS and minimum initial dose
    using the pulse frequency 
    and dose rate indicated. 
    The temperature assigned to the XPCS diffusion constants is estimated
    based on the respective temperature rise shown in \textbf{a}. 
    The base temperature was $T_0=\SI{298}{\kelvin}$ for all XPCS measurements. Source data are provided as a Source Data file.
    }\label{fig:temperature-dose-rate}
\end{figure}

The increase of temperature for the different XFEL parameters is estimated 
following the model used by \textcite{lehmkuhler_emergence_2020} using a
weighted average heat capacity of the constituents of
$c_p=\SI{3.42}{\joule\per\gram\per\kelvin}$.
With a maximum number of \num{3.82e8}~photons per X-ray pulse (\Cref{tab:beamline-params})
the energy density is \SI{5.5}{\mJ\per\mm\squared}, which is an order of magnitude smaller
than in the work of \textcite{lehmkuhler_emergence_2020}.
\Cref{fig:temperature-dose-rate}a displays the corresponding temperature rise  
during a pulse train with a temperature increase of 
$\Delta T  \approx \SI{8}{\kelvin}$ after the low intensity pulse trains
(blue and orange) 
and $\Delta T \approx \SIrange{19}{21}{\kelvin}$ after the high intensity
pulse trains (red and green).
For comparison, we measured the temperature dependence of the equilibrium 
diffusion constant with dynamic light scattering~(DLS), where the sample was 
equilibrated at each temperature before a measurements,
and display the results in \Cref{fig:temperature-dose-rate}b. 
The values of $D_0$ measured with MHz-XPCS and 
X-ray pulse repetition rates of \SI{1.13}{\mega\hertz} are close to their equilibrium
values obtained from DLS, when taking the XFEL-induced temperature increase
into account. In contrast, employing higher 
XFEL frequencies of \SI{2.26}{\mega\hertz} yields consistently higher diffusion
coefficients which cannot be explained by a temperature increase alone. 
We hypothesize that the intense MHz XFEL pulses create a non-equilibrium 
state triggering processes on the sub-microsecond time scale.
One example of such processes is the spatial homogenization of the 
aforementioned radiolysis products.
The typical rates of secondary products such as \ce{OH-} radicals are on the order
of microseconds \cite{attri_generation_2015,baba_quantitative_2021}. 
Thus, on sub-microsecond time scales, the XFEL pulses simultaneously produce
and probe a spatially inhomogeneous local distribution of the radiolysis products. 
The resulting chemical gradients, molecular repulsion due to dose rate dependent
protein charging, and possibly changes of the ionic strength of the solution, as well as damage to the PEG molecules, 
could contribute to the observed enhanced diffusive motion. Clearly, more 
systematic data and additional work by theory and simulation is needed to
understand this XFEL driven motion.  

\begin{figure}[H]
  \centering
  \includegraphics[width=\linewidth, trim={0 4mm 0 0}]{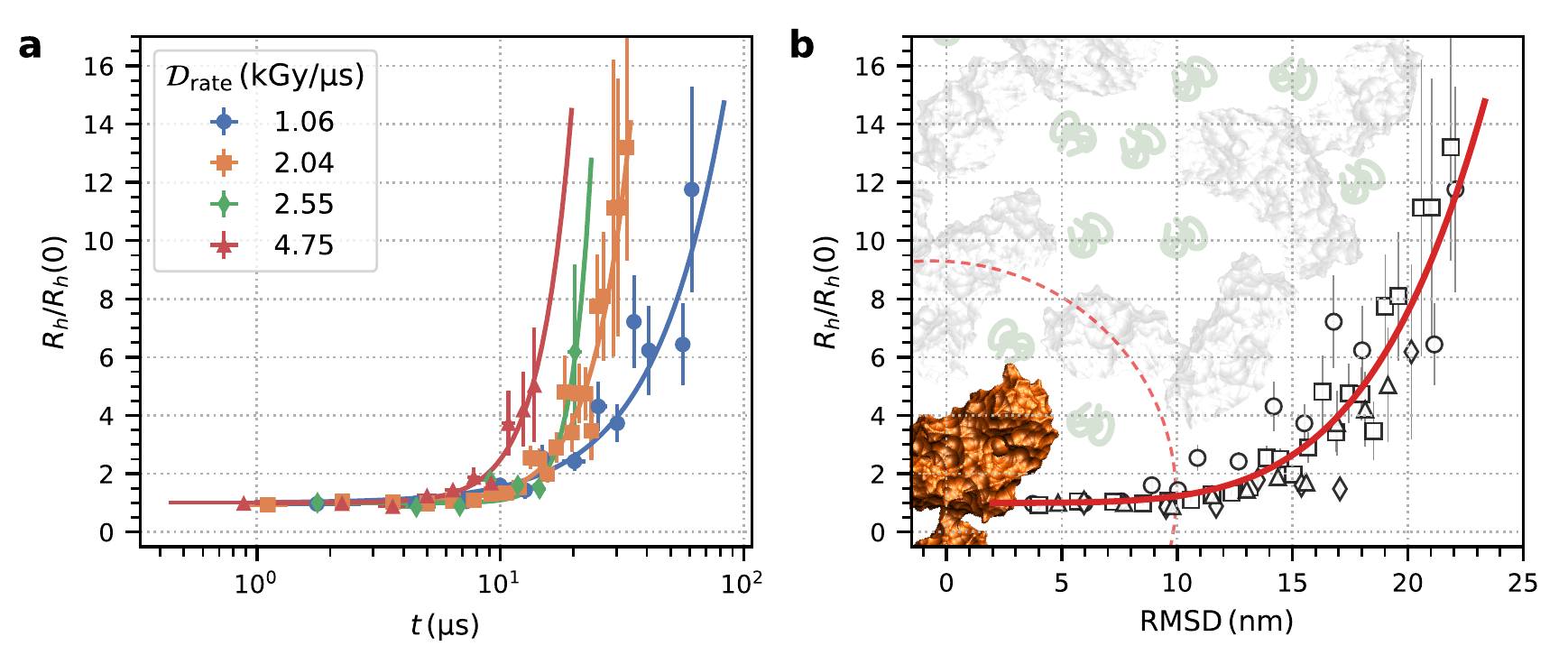}
  \caption[Rhyd and msd]{\textbf{X-ray induced aggregation pathways.} 
  \textbf{a}~Apparent hydrodynamic radii normalized to the initial value $R_h(0)$
  as a function of measurement time for different dose rates.
  The solid lines are guides to the eye. 
  \textbf{b}~Apparent hydrodynamic radii normalized to the initial value $R_h(0)$
  as a function of root mean square displacement (RMSD).
  The red solid line is a guide to the eye. 
  The dashed red circle describes a sphere with a radius of about \SI{10}{\nm}
  and marks the space an Ig molecule can explore before the onset of aggregation. Source data are provided as a Source Data file.
  }\label{fig:rmsd}
\end{figure}

The question arises why the faster dynamics at higher dose rates do not lead to a
dose rate dependent aggregation visible in the SAXS signal.
We address this question by employing the Stokes-Einstein relation and estimating
the temporal evolution of the relative changes to the 
apparent hydrodynamic radii via $R_h(t)/R_h(0) = D_0(0)/D_0(t)$,
where $D_0(0)$ and $R_h(0)$ represent the 
respective values at the minimum dose in  \Cref{fig:diffusion-coeff-kww-exponent}.
The increase of this ratio serves as an indicator for protein aggregation. 
\Cref{fig:rmsd} shows that aggregation sets in earlier and develops faster for 
higher dose rates.
For a dose rate of \SI{1.06}{\kilo\gray\per\micro\second},
$R_h(t)/R_h(0)$ has approximately doubled after  \SI{18.0(9)}{\micro\second} and after
\SI{6.8(6)}{\micro\second} for \SI{4.75}{\kilo\gray\per\micro\second} 
(Supplementary Fig.~4).
Using the measured diffusion coefficients $D_0$, we further calculate the time
dependent root mean square displacement~(RMSD) $\sqrt{\langle \Delta 
x^2(t)\rangle} = \sqrt{6D_0t}$
of the proteins and plot $R_h(t)/R_h(0)$ as a function of RMSD 
(\Cref{fig:rmsd}b).
The data for the different dose rates collapse
onto a single master curve (red line) indicating that the
onset of aggregation mainly depends on the RMSD of the protein molecules.
Higher dose rates induce faster movement of the proteins, and thus the RMSD necessary 
for aggregation is reached earlier.
\Cref{fig:rmsd}b also reveals that aggregation sets in after a RMSD 
of about \SI{10}{\nm} and the space a single protein can explore in the crowded 
solution before that happens is indicated by a the red dashed circle.
This area is quite large considering that the sample is a densely packed solution of
\SI{250}{\mg\per\milli\litre}, where the mean free path $l$ between two molecules is
typically smaller than their radius.
We estimate $l=1 / (n\pi (2R_h)^2) = \SI{2.6}{\nano\metre}$ from 
the number density $n$ and the molecular radius $R_h=\SI{5.5}{\nano\metre}$
of an Ig molecule which in turn  implies an average number of contacts between
proteins on the order of  $N=\langle \Delta x^2(t)\rangle/l^2 \approx 14$
before aggregation sets in.

Our analysis indicates that aggregation is not strictly translational 
diffusion limited, but multiple contacts are necessary to attach two protein 
molecules to each other and form aggregates.
This may hint towards the importance of specific interaction sites driving the 
aggregation process \cite{northrup_kinetics_1992}.
In addition, we note that unfolding processes which increase the protein 
propensity to aggregate do also occur on time scales of microseconds 
\cite{kubelka_protein_2004}.
Thus, the observed initial period of constant $R_h$ points towards a minimum 
incubation time on the order of \SI{10}{\micro\second} before the proteins
locally unfold or a time needed for rotational motion of molecules in order to allow activated sites to form local bonds.
This incubation time and the minimum RMSD of \SI{10}{\nm} define a window of 
opportunity where dynamics can be measured in a \emph{correlation before aggregation} scheme.
Analogously to \emph{diffraction before destruction} on femtosecond time scales~\cite{nass_radiation_2019,nass_indications_2015,schlichting_serial_2015,neutze_potential_2000},
\emph{correlation before aggregation} will allow to obtain experimental
information about the structural sample dynamics on (sub-)microsecond time 
scales before X-ray induced changes will become apparent in the XPCS 
signal.
Additional data with more dose rates could allow developing novel methods to
estimate the diffusion coefficients at zero dose rate
\cite{chushkin_deciphering_2020}. However, the dose rate dependence might be highly 
dependent on the sample. On the other hand, developing the 
\emph{correlation before aggregation} approach further would provide a window 
of opportunity, where with moderate doses and dose rates, the SNR is increased
and the overall measurement time and sample consumption could be reduced.
Reducing the sample consumption is crucial for measuring particularly precious solutions
or systems that exhibit phase transitions on microsecond time scales, e.g., biomolecular 
condensates, which is hard to repeat thousands of times 
\cite{girelli_microscopic_2021,begam_kinetics_2021,ragulskaya_interplay_2021}.
Improving the experiments could
be achieved for instance by making use of the self-seeding schemes
which provide a much larger longitudinal coherence length. A larger longitudinal coherence
length would allow for a larger beam size with similar measured speckle contrast yielding
a lower photon density on the sample.
This reduces the radiation damage to the sample and the amount of sample needed. 
It also increases the scattering volume and scattering intensity and thus
strongly increases the signal-to-noise ratio~\cite{moller_x-ray_2019}.
Further technical improvements such as MHz detectors with smaller pixel
size are needed to improve the SNR even further which allows extending 
the accessible $q$-range and lowering the dose rate needed.

Summarizing, we demonstrated that MHz-XPCS bears the potential to become a
useful tool for measuring dynamics of biological macromolecules in solution 
on molecular length scales and on the time scales relevant 
for diffusive motion in cells. 
Importantly, our results indicate that taking the temperature 
rise of the solution into account allows for studying equilibrium dynamics
within the first coordination shell of the molecules.
Higher XFEL frequencies drive the dynamics and lead to increasing diffusion
coefficients and aggregation which sets in after a time window of \SI{10}{\micro\second}. We refer to this approach as
\emph{correlation before aggregation} which allows to capture protein dynamics
in solution before the manifestation of X-ray induced effects.
Additional experiments and simulations are needed to fully understand 
the underlying physics of the involved processes.
Understanding the observed dose rate dependence of the diffusion process involves
accurate knowledge of a number of yet unknown factors,
such as the role of the interaction potentials, concentration, 
solvent chemical composition, and size and masses of the proteins.
Resolving these properties and the role of radiolysis processes and their products
in this context will determine the best data acquisition strategies for measuring
the unperturbed dynamic properties.

\section*{Methods}

\subsection*{Sample Preparation}
The sample preparation followed a procedure provided by the
literature \cite{da_vela_effective_2017}. Polyclonal bovine immunoglobulin (purity 
$\geq 99\%$, Sigma-Aldrich, SRE0011), PEG 1000 (Sigma-Aldrich, 81188), NaCl (Merck
106404), HEPES (Roth, HN78) and NaN$_3$ (Sigma-Aldrich, S8032) were used as
received. All solutions were prepared in a buffer of composition 20 mM HEPES pH
= 7.0, 2~mM NaN$_3$, using degassed Milli-Q water (Merck Millipore 18.2~M$\Omega\cdot$cm).
The concentration of the immunoglobulin stock solutions
was assessed by UV absorption at 280nm, using an extinction coefficient of
$e_{280}=\SI{1.4}{\milli\litre\per\milli\gram\per\centi\metre}$ with a Cary 50 UV-Vis
spectrophotometer. The experimental phase diagram of this system has been
established in our previous work \cite{da_vela_effective_2017}. The ``parent
solution'' was equilibrated for about 24 h at \SI{294}{\kelvin} and then briefly
centrifuged, resulting in a clear dense and a dilute phase, separated by a sharp
meniscus. The parent solution composition was immunoglobulin 200~mg/mL, PEG
12\% w/v and NaCl 150~mM. The dense liquid phase was used for XPCS measurements 
with a concentration of roughly \SI{250}{\milli\gram\per\milli\liter}.

\subsection*{Measurement Protocol}
The sample was moved continuously through the X-ray beam to refresh the sample
volume between trains. The translation motor speed was \SI{0.4}{\mm\per\s} 
leading to an absolute sample translation of \SI{40}{\micro\m} between 
successive trains (a train arrives every \SI{100}{\ms}). X-ray generated heat 
diffuses with a thermal diffusion constant of 
$D_T=\SI{0.143}{\mm\squared\per\s}$. Assuming homogeneous heat diffusion in 
radial direction perpendicular from the beam, after $\tau_\mathrm{train}\SI{100}{\ms}$ the heat 
diffused in a cylinder with radius $\sqrt{6D_T\tau_\mathrm{train}}\approx\SI{293}{\micro\m}$.
The illuminated volume of the next train is then 
$\SI{10}{\micro\m\squared}/\SI{293}{\micro\m\squared}\approx\SI{0.1}{\percent}$ of that volume.
Therefore, beam-induced heating effects generated by the previous train can be neglected for the following illumination.

\subsection*{Calculation of the Absorbed Dose and Dose Rate}
On the time scale of an individual X-ray pulse ($\leq\SI{50}{\fs}$~\cite{madsen_materials_2021}) and with maximum 
flux, the peak dose rate can reach several hundred mega Gray per pulse~\cite{chapman_femtosecond_2011}.
Utilizing MHz repetition rates and high attenuation, 
the X-ray pulses are delivered on (sub-)~microsecond time scales
such that an average dose rate on the order of several kilo Gray per microsecond can be reached.
Correspondingly, synchrotron sources produce typical average dose rates of kilo Gray per second. 

In order to quantify the amount of energy absorbed by a certain sample mass, we
calculate the dose, $\mathcal{D}(N_p)$, absorbed by the sample after $N_p$ pulses:
\begin{equation}
  \label{eq:radiation-dose}
  \mathcal{D}(N_p) = \bigg\langle\sum_{i=1}^{N_p}\frac{\Phi_c^j(i) \, E_c\, A}
  {z^2 d_s\,\rho}\bigg\rangle_{\mathrm{j}}\,,
\end{equation}
where $A=(1-e^{-d_s/\mu_\mathrm{eff}})$ denotes the sample absorption where ${d_s=\SI{1.5}{\mm}}$ is the 
sample thickness and $\mu_\mathrm{eff}=\SI{1.35}{\mm}$ is the effective attenuation length of the solution 
calculated as the weighted harmonic mean of the individual 
components~\cite{seltzer_tables_1995,girelli_microscopic_2021},
${E_c=\SI{9}{\kilo\electronvolt}}$ the photon energy, $\Phi_c^j(i)$ the number of photons in 
pulse $i$ in train $j$, $\rho=\SI{1.09}{\g\per\cubic\cm}$ is the sample 
density, and $z=\SI{10}{\micro\metre}$ is the beam size. $\langle \dots \rangle_{\mathrm{j}}$ 
denotes an average over trains. The dose is measured in
\emph{Gray}~($\SI{1}{\gray}=\SI{1}{\joule\per\kilo\gram}$).

To account for SASE related intensity fluctuations, the average dose rate is calculated as the average dose absorbed by the sample after ten
pulses divided by the delay between two successive pulses, $\tau_p$:
\begin{equation}
  \label{eq:avr-dose-rate}
  \mathcal{D}_{\mathrm{rate}} = \frac{\mathcal{D}(N_p=10)}{10\tau_p} \,.
\end{equation}

\subsection*{Error Bar Calculation}
The error bars of the correlation functions in \Cref{fig:correlation-functions}
are calculated as the standard deviation of the fluctuations
of the contrast values within
a $q$-bin. This standard deviation is used to calculate the 
weighted average over trains and times in 
\Cref{eq:two-time-correlation,eq:time-resolved-correlation}. 
These error bars are used in the fits for the parameters estimation. 
The error bars in the following plots, which display the fit results of 
the correlation functions, indicate the parameter uncertainty obtained
from the fits using least-squares minimization.

\subsection*{X-ray Induced Heating}
The relaxation time of heat diffusion used in \Cref{fig:temperature-dose-rate} 
to estimate the X-ray induced heating, is calculated as 
$\tau_\mathrm{heat}= c_p \rho  z^2 / (2 k_w)=\SI{310}{\micro\second}$, where 
$c_p=\SI{3.42}{\joule\per\gram\per\kelvin}$
is the heat capacity of the solution and $k_w=\SI{0.6}{\watt\per\metre\per\kelvin}$ 
is the thermal conductivity. The heat capacity is calculated by an average of the heat capacities of water,
PEG, and IgG weighted with their volume fractions.
The temperature increase induced by a pulse with energy $E_p$ is 
$\Delta T_\mathrm{max} = 4\log(2)E_p / (2\pi c_p \rho z^2  \mu_\mathrm{eff})$.
This formalism leads to values from
$\Delta T=\SIrange{0}{21}{\kelvin}$ depending on $E_p$ and the number of X-ray
pulses (\Cref{fig:temperature-dose-rate}). It should be noted that  
\SI{1}{\percent} and less of the maximum possible $E_p$ at MID has been used 
to reduce beam-induced effects.

\subsection*{Data availability}
The data are available from the authors upon request.

\subsection*{Code availability}
Analysis scripts are available from the authors upon request.

\printbibliography

\section*{Acknowledgements}
We acknowledge the European XFEL in Schenefeld, Germany, for provision of X-ray free
electron laser beamtime at Scientific Instrument MID (Materials Imaging and
Dynamics) and would like to thank the staff for their assistance. We acknowledge
DESY (Hamburg, Germany), a member of the Helmholtz Association HGF, for the
provision of experimental facilities. This research is supported by Center of Molecular Water Science
(CMWS) of DESY in an Early Science Project, the MaxWater initiative of the
Max-Planck-Gesellschaft and the Wenner-Gren Foundations. This work is supported by the
Swedish National Research Council Vetenskapsr\text{\aa}det (
2019-05542, FP), the R\"ontgen-\text{\AA}ngstr\"om Cluster ( 
2019-06075 FP), BMBF (05K19PS1 and 05K20PSA, CG ; 05K19VTB, FS and FZ), DFG-ANR (SCHR700/28-1,
SCHR700/42-1, ANR-16-CE92-0009, FS and FZ).
AR acknowledges the Studienstiftung des deutschen Volkes for a personal fellowship.
NB acknowledges the Alexander von Humboldt-Stiftung for postdoctoral research fellowship.

\section*{Author contributions}

FP, FZ, FS, and CG conceived the experiment, which was designed and coordinated by MR,
FP, FZ and CG. AG and AR prepared and handled the samples. UB, JH, JM, ARF, MS,
RS, AZ, AM operated MID and collected data together with MR, AG and AR. MR, AG, AR,
FP, ST, SD, MSA and MB performed online data processing and analysis, whereas SB,
HFP, MLP, MF, and LR were in addition responsible for the elog. MR performed offline data
processing and analysis. MR, FP, FS, FZ, and CG discussed the XPCS data analysis
with input from all authors. 
R.R and R.S supported the development of analysis routines and the
integration of experimental equipment. All authors jointly performed the
experiment and discussed the final results, to a large extent remotely due to
travel restrictions imposed by the COVID-19 pandemic. The manuscript was written
by MR, FP and CG with input from all authors.

\section*{Competing interests}
The authors declare no competing interests.

\begin{table}[H]
  \centering
  \caption{\textbf{Measurement parameters:} $\mathcal{D_{\mathrm{rate}}}$ is the average dose
    rate, $f_{\mathrm{FEL}}$ is the XFEL frequency, and $\tau_p$ is the time between successive pulses
    which defines the the minimum XPCS delay time. $T_{\mathrm{cvd}}$ is the transmission of the diamond 
    attenuators which results in the average number of incident photons per
    X-ray pulse (ph/pls) on the sample, $\langle\Phi_c\rangle$.
    $N_{\mathrm{train}}$ is the number of pulse trains averaged in the
    analysis. $N_p$ is the maximum number of pulses per train.}\label{tab:beamline-params}
  \begin{tabular}{
    S[table-format=1.2,table-column-width=1.5cm]
    S[table-format=1.1,table-column-width=1cm]
    S[table-format=1.0,table-column-width=1cm]
    S[table-format=1.1,table-column-width=1cm]
    S[table-format=1.2,table-column-width=2cm]
    S[table-format=3.0,table-column-width=1.1cm]
    S[table-format=2.0,table-column-width=1cm]
    }
    \toprule
    {\multirow{2}{1.5cm}{\(\mathcal{D}_{\mathrm{rate}}\) \\ \((\si{\kilo\gray\per\micro\second})\)}}
    & {\multirow{2}{1cm}{\( f_{\mathrm{FEL}}\) \\ \( (\si{\mega\hertz}) \) }}
    & {\multirow{2}{1cm}{\(\tau_{p}\) \((\si{\nano\s})\)}}
    & {\multirow{2}{1cm}{\( T_\mathrm{cvd}\) \\ \((\si{\percent}) \) }}
    & {\multirow{2}{2cm}{\( \langle\Phi_c\rangle\) \\ \((10^8\si{ph/pls}) \) }}
    & {\multirow{2}{1.1cm}{\( N_{\mathrm{train}} \) }}
    & {\multirow{2}{1cm}{\( N_p \) }} \\ \\
    \midrule
    1.06 & 1.13 & 886 & 0.6 & 1.59 & 2800 & 144 \\
    2.04 & 2.26 & 443 & 0.6 & 1.53 & 9200 & 200 \\
    2.55 & 1.13 & 886 & 1.4 & 3.82 & 2000 & 144 \\
    4.75 & 2.26 & 443 & 1.4 & 3.56 & 5200 & 200 \\
    \bottomrule
  \end{tabular}
\end{table}

\end{document}


\author[1]{Mario Reiser\thanks{mario.reiser@fysik.su.se}}
\author[2]{Anita Girelli}
\author[2]{Anastasia Ragulskaya}
\author[1]{Sudipta Das}
\author[1]{Sharon Berkowicz}
\author[1]{Maddalena Bin}
\author[1]{Marjorie Ladd-Parada}
\author[1]{Mariia Filianina}
\author[1,2]{Hanna-Friederike Poggemann}
\author[2]{Nafisa Begam}
\author[3]{Mohammad Sayed Akhundzadeh}
\author[3]{Sonja Timmermann}
\author[3]{Lisa Randolph}
\author[4]{Yuriy Chushkin}
\author[5]{Tilo Seydel}
\author[6]{Ulrike Boesenberg}
\author[6]{J\"org Hallmann}
\author[6]{Johannes M\"oller}
\author[6]{Angel Rodriguez-Fernandez}
\author[6]{Robert Rosca}
\author[6]{Robert Schaffer}
\author[6]{Markus Scholz}
\author[6]{Roman Shayduk}
\author[6]{Alexey Zozulya}
\author[6]{Anders Madsen}
\author[2]{Frank Schreiber}
\author[2]{Fajun Zhang}
\author[1]{Fivos Perakis\thanks{f.perakis@fysik.su.se}}
\author[3]{Christian Gutt\thanks{christian.gutt@uni-siegen.de}}

\affil[1]{Department of Physics, AlbaNova University Center, Stockholm University, S-106 91 Stockholm, Sweden}
\affil[2]{Institut f\"ur Angewandte Physik, Universit\"at T\"ubingen, Auf der Morgenstelle 10, 72076 T\"ubingen, Germany}  
\affil[3]{Department Physik, Universit\"at Siegen, Walter-Flex-Strasse 3, 57072 Siegen, Germany}
\affil[4]{The European Synchrotron, 71 Avenue des Martyrs, Grenoble, 38000, France}
\affil[5]{Institut Laue-Langevin, 38042 Grenoble Cedex 9, France}
\affil[6]{European X-Ray Free-Electron Laser Facility, Holzkoppel 4,22869 Schenefeld, Germany}

\maketitle

\section*{Coherence and Speckle Contrast}

The experimental speckle contrast, $\beta_\mathrm{exp}$ depends on nearly all experimental parameters such as pixel size 
speckle size, beam size, sample thickness, momentum transfer $q$, the transverse
and longitudinal coherence properties of the X-rays, etc. $\beta_\mathrm{exp}$ can be calculated 
as the product of the longitudinal contrast factor, $\beta_l$, and the transverse contrast factor, $\beta_t$:
\begin{equation}
    \label{eq:speckle-contrast-bare}
    \beta_\mathrm{exp}=\beta_l\beta_t\,.
\end{equation}

\begin{figure}[htbp]
  \centering
  \includegraphics[trim={0 4mm 0 0}]{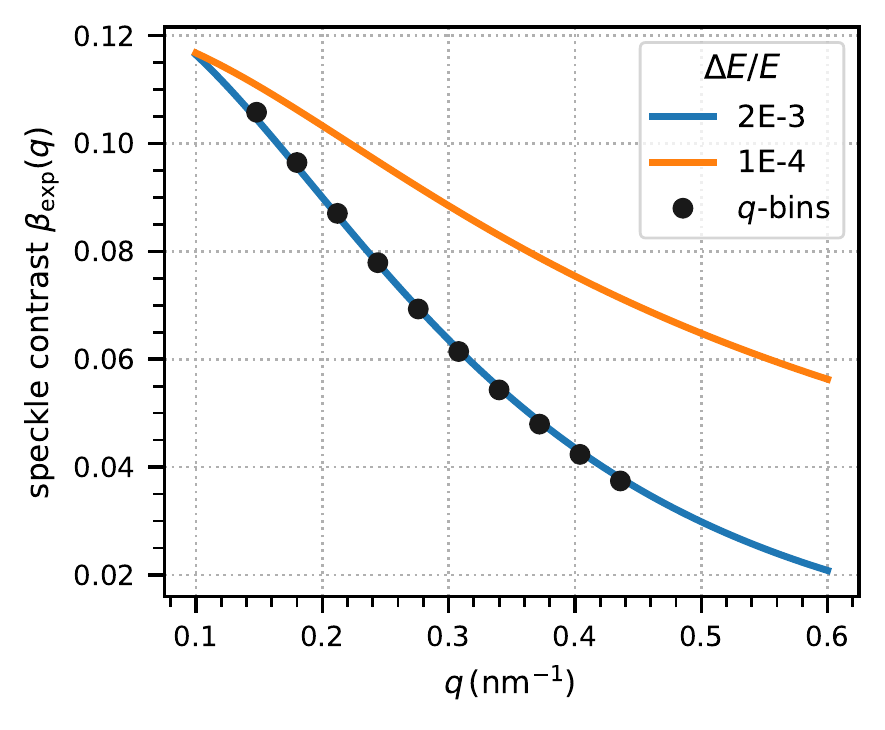}
  \caption[speckle-contrast]{\textbf{Speckle contrast for different energy bandwidths}. $\Delta E/E=\num{2e-3}$ corresponds to the SASE bandwidth and  $\Delta E/E=\num{1e-4}$ to a monochromatic beam. The black points indicate the contrast values in the $q$-bins used
  to fit the experimental data.} \label{fig:speckle-contrast}
\end{figure}

For XFELs, the model described in \textcite{hruszkewycz_high_2012} is often employed to estimate $\beta_l(q)$. 
A detailed  description of the mathematical formalism can be found in the supplementary material of \cite{hruszkewycz_high_2012}.
\textcite{lehmkuhler_emergence_2020} show that the speckle contrast at EuXFEL can
be described by this model as well. 
$\beta_l$ is determined by the energy bandwidth, $\Delta E/E$, which can be decreased by using a seeded beam or a monochromator. 
Both were not available for this experiment. Instead, the pink SASE beam was used with an energy bandwidth of $\Delta E/E\approx \num{2e-3}$.
For the transverse coherence factor, $\beta_t\approx 0.5$ was found for different XFELs including European XFEL
\cite{madsen_materials_2021,lehmkuhler_emergence_2020,lehmkuhler_single_2014,alonso-mori_x-ray_2015}.
Eventually, the following model was used to describe the data
\begin{equation}
    \label{eq:speckle-contrast-exp}
    \beta_\mathrm{exp}(q)=0.5\,\beta_l(q)\,.
\end{equation}
\Cref{fig:speckle-contrast} displays $\beta_\mathrm{exp}(q)$ as calculated by \Cref{eq:speckle-contrast-exp} as function of the momentum transfer assuming a beam size of \SI{10}{\micro\metre}, a sample-detector distance of \SI{7.46}{\m}, a photon energy of \SI{9}{\kilo\electronvolt}, a sample thickness of \SI{1.5}{\mm}, a pixel size of \SI{200}{\micro\metre}, and a bandwidth of \num{2e-3} and \num{1e-4}, respectively. 
The blue line indicates the contrast during the experiment where the 
black dots are the corresponding values used to model the correlation functions.
The orange line shows the increased  speckle contrast when using a smaller bandwidth.

%
%

\section*{Signal-to-Noise Ratio}

Following the work of \textcite{falus_optimizing_2006},
the XPCS signal-to-noise ratio,  $R_{sn}$, can be calculated as
\begin{equation}
  \label{eq:snr}
  R_{sn} = \beta I \sqrt{N_{p}N_\mathrm{trains}N_\mathrm{pix}}\,,
\end{equation}
where $\beta$ is the speckle contrast, $I$ is the intensity in numbers of photons 
per pixel, $N_{p}=20$ is the number of pulses or images 
used to calculate the first point of the correlation functions, 
$N_\mathrm{trains}$ is the number of trains that are averaged, 
$N_\mathrm{pix}=7494$ is the number of pixels in the $q$-bin where 
the correlation function is calculated.

\begin{figure}[htbp]
  \centering
  \includegraphics[trim={0 4mm 0 0}]{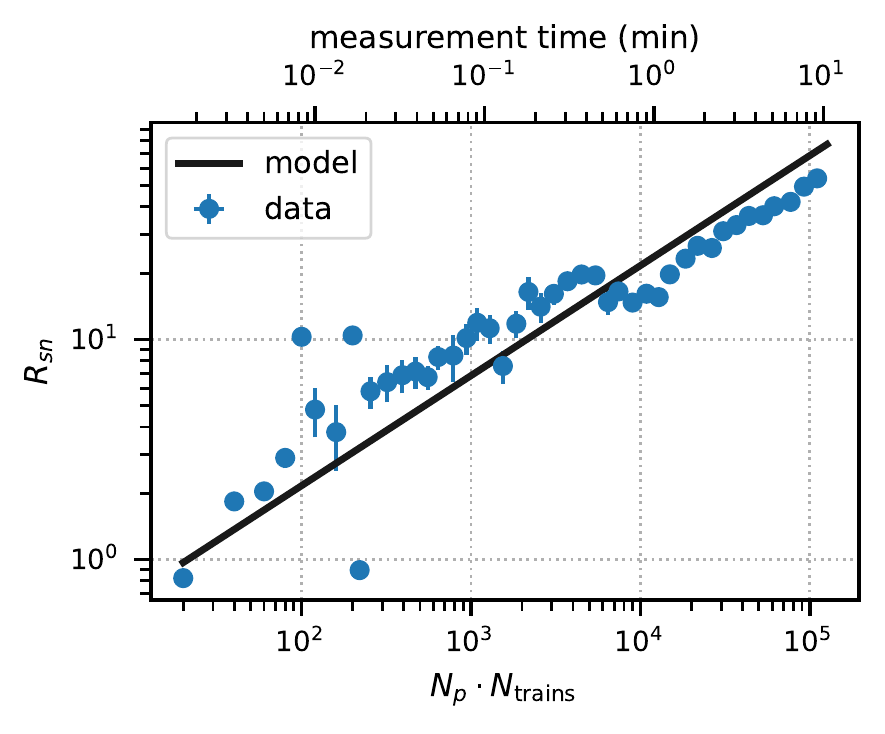}
  \caption[signal-to-noise-ratio]{\textbf{Signal-to-noise ratio as a function of the number of X-ray pulses.} The signal-to-noise ratio of the experimental data compared with the estimations of \Cref{eq:snr}. On the top, the total measurement time is indicated assuming that every train measured is suitable for the analysis.} 
  \label{fig:signal-to-noise-ratio}
\end{figure}

\Cref{fig:signal-to-noise-ratio} displays $R_{sn}$ of the first point of the correlation functions at \SI{0.148}{\per\nm} as a function of the total number of X-ray pulses, where each of the $N_\mathrm{trains}$ illuminated a fresh sample volume. The data have been rebinned along the abscissa and the error bars indicate the standard deviation within each bin. In addition, the total measurement time is indicated assuming that every pulse is used for the analysis. The measurement time obviously  depends on both machine performance and filtering criteria applied during the  experiment as not every train might be usable for the analysis, e.g., because of very low intensity. In \Cref{fig:signal-to-noise-ratio}, it is assumed that every train is used for the analysis. In our experiment  about \SI{20}{\percent} of the trains were discarded. The visible fluctuations in $R_{sn}$ are probably related to fluctuating instrument parameters and varying machine performance. 

The primary advantage of using a monochromatic beam is that it would allow for a larger beam size with similar
contrast, hence yielding a lower photon density on the sample. This reduces the radiation damage to the sample
and the amount of sample needed. It also increases the scattering volume and scattering intensity and thus
strongly increases the signal-to-noise ratio~\cite{moller_x-ray_2019}.

\section*{Azimuthal Dependence of the Scattering Signal}
\begin{figure}[htbp]
  \centering
  \includegraphics[trim={0 4mm 0 0}]{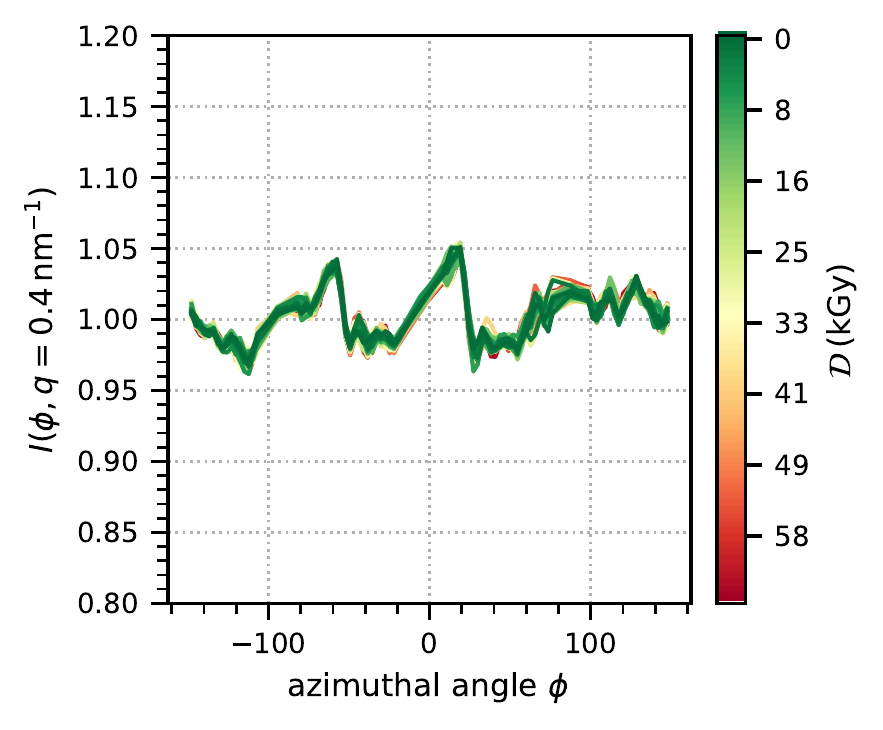}
  \caption[anisotropy-scattering-signal]{\textbf{Azimuthal scattering intensity.} Integrated scattering signal at 
  $q=\SI{0.4}{\per\nm}$ as a function of azimuthal angle $\phi$.
  $\phi=0$ corresponds to the horizontal direction. 
  The data have been normalized to the average value of each curve. The color indicates the absorbed dose. The lines of different doses are overlapping almost perfectly.} \label{fig:anisotropy-scattering-signal}
\end{figure}

\Cref{fig:anisotropy-scattering-signal} displays the scattering signal from Fig.\,$2$ in the
main manuscript at $q=\SI{0.4}{\per\nm}$ as a function of the azimuthal angle $\phi$ for
different doses.
The data have been normalized to the average value 
of each curve to exclude any effect on the average scattering signal. The small kinks can be 
attributed to the AGIPD detector. The signal does not exhibit any sign of anisotropy within 
the measurement accuracy. Furthermore, the shape of the curves does not change as a function 
of absorbed dose indicating that the signal stays isotropic throughout the measurements. 

\section*{Dose dependence of the Hydrodynamic Radius}
\begin{figure}[htbp]
  \centering
  \includegraphics[scale=.8,trim={0 4mm 0 0}]{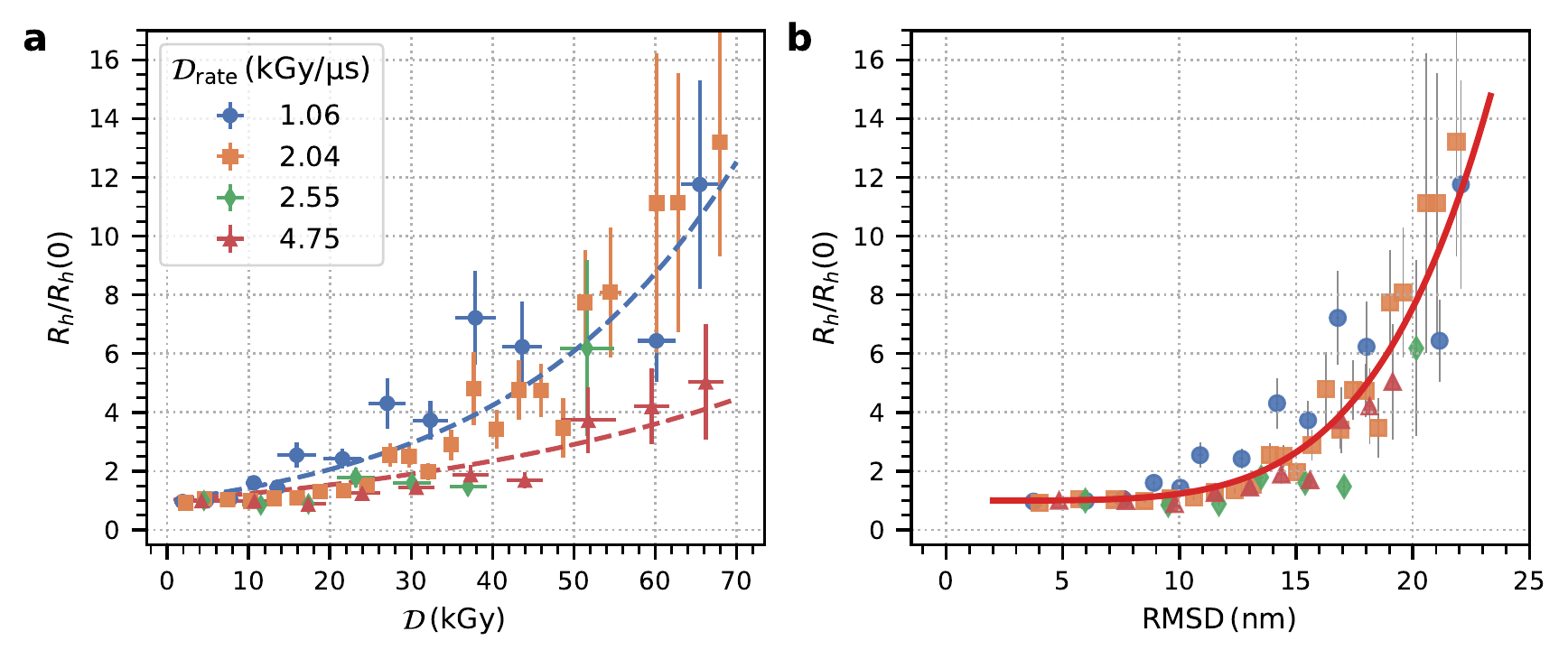}
  \caption[rhyd-msd-dose]{\textbf{Dose dependence of the hydrodynamic radius.} (\textbf{a}) Normalized hydrodynamic radius $R_h$ as function of dose. (\textbf{b})  
  Normalized hydrodynamic radius $R_h$ as function of root mean squared displacement. The dashed lines in (\textbf{a}) indicate fits with an exponential model. The error bars describe the fit accuracy.} \label{fig:hydrodynamic-radius-vs-dose}
\end{figure}

We plot the normalized hydrodynamic radius $R_h$ as a function of dose in 
\Cref{fig:hydrodynamic-radius-vs-dose}a and as a function of root mean squared displacement  in \Cref{fig:hydrodynamic-radius-vs-dose}b for comparison as in the main manuscript. From fitting the data of the 
lowest and highest dose rate (dashed lines) with an exponential model we conclude that $R_h$ increases by a factor of two after
\SI{19.2(10)}{\kilo\gray} for a dose rate of \SI{1.06}{\kilo\gray\per\micro\second} and after \SI{32(3)}{\kilo\gray} 
for a dose rate of \SI{4.75}{\kilo\gray\per\micro\second}. These values correspond to 
starting times of \SI{18.1(9)}{\micro\second} and \SI{6.8(6)}{\micro\second}, 
respectively. 

\newpage
\printbibliography